\documentclass[preprint,showpacs,preprintnumbers,amsmath,amssymb]{revtex4}

\usepackage{amsmath}
\usepackage{graphicx}
\usepackage{dcolumn}
\usepackage{bm}

\begin{document}

\title{Electrical rectification effect in single domain magnetic 
microstrips: a micromagnetics-based analysis}

\author{Andr{\'{e}} Thiaville}
\affiliation{Laboratoire de Physique des Solides, CNRS,
Univ. Paris-sud, 91405 Orsay Cedex, France}
\author{Yoshinobu Nakatani}
\affiliation{Department of Computer Science, University of 
Electrocommunications, Chofu, Tokyo 182-8585, Japan}


\begin{abstract}
Upon passing an a.c. electrical current along magnetic micro- or nanostrips, 
the measurement of a d.c.
voltage that depends sensitively on current frequency and applied field has been
recently reported by A. Yamaguchi and coworkers.
It was attributed to the excitation of spin waves by the spin transfer torque, 
leading to a time-varying anisotropic magnetoresistance and, by mixing of a.c. 
current and resistance, to a d.c. voltage.
We have performed a quantitative analysis by micromagnetics, including 
the spin transfer torque terms considered usually, of this situation.
The signals found from the spin transfer torque effect are several orders of 
magnitude below the experimental values, even if a static inhomogeneity of
magnetization (the so-called ripple) is taken into account.
On the other hand, the presence of a small non-zero average {\OE}rsted field
is shown to be consistent with the full set of experimental results, both
qualitatively and quantitatively.
We examine, quantitatively, several sources for this average field and
point to the contacts to the sample as a likely origin.
\end{abstract}

\pacs{72.25.Pn, 76.50.+g, 41.20.-q, 72.15.-v}
\maketitle

\section{Introduction}
\label{sec:intro}

The possibility to act on the magnetization of a sample by an electrical 
current
within it, not through the classical {\OE}rsted field but through the 
spin-polarization of electrical current in ferromagnets, offers fascinating
opportunities in nanomagnetism and nanoelectronics
\cite{Berger96,Slonczewski96}.
In the situation where the sample consists of separated and 
uniformly magnetized media crossed by the current, the description of 
the physics appears simpler and, indeed, agreement between experiments and 
modelling does not appear out of reach
\cite{Krivorotov05,Berkov08}.
However, when the current flows in a magnetic medium with a continuously 
varying magnetization, the situation is more complex. 
As a result, several forms for this so-called spin transfer torque (STT) 
have been proposed \cite{Bazaliy98,Thiaville04,Zhang04,Thiaville05}, and the 
appropriate equation for magnetization dynamics has even been questionned 
\cite{Stiles07,Smith07}.

In such a situation, the more experimental results in different configurations 
is clearly the better. 
Among these, the recent discovery of an electrical rectification effect in
magnetic strips with widths of the order of a micrometer and thicknesses of the
order of a few tens of nanometers \cite{Yamaguchi07} is especially appealing.
The effect was observed for current densities below or of the order of those 
required for STT to act on domain walls.
However, a relatively large static field was applied so that the strip was 
in a single domain state, contrarily to the situation where a signal was
measured in presence of a domain wall \cite{Bedau07,Moriya08}.
This last feature is puzzling.
Indeed, STT within a continuous magnetization structure is only expected when
a magnetization gradient exists.
In the simplest STT formulation, valid for slow magnetization variations with
respect to electrons' spin precession or diffusion length, the STT is namely 
expressed as
\begin{equation}
\label{eq:STT}
 \frac{\partial \vec{m}}{\partial t}|_{STT} =
-\left( \vec{u}\cdot \vec{\nabla} \right) \vec{m}
+ \beta \vec{m} \times \left[\left( \vec{u}\cdot \vec{\nabla} \right) 
\vec{m} \right],
\end{equation}
where the velocity $\vec{u}$ is an expression of the current density $\vec{J}$
with spin polarization $P$ according to
\begin{equation}
\label{eq:u}
\vec{u} = \vec{J} \, \frac{g \mu_\mathrm{B} P}{2 e M_\mathrm{s}}.
\end{equation}
The (small) number $\beta$ has been related to spin flip of the 
conduction electrons, in 
several models \cite{Zhang04,Thiaville05,Tatara07,Piechon07}.
From (\ref{eq:STT}), one sees that magnetization gradients along the electric 
field are required.
Such gradients should however not exist in the experimental situation 
considered above (long strip under a large field), at least for perfect samples.
The possibility mentioned by the authors is that the uniform state becomes 
unstable under a.c. current at an appropriate frequency, as indeed predicted 
for very large d.c. currents \cite{Shibata05}.

The object of this paper is to perform a full micromagnetic analysis 
of the situation in order to analyze the various sources of rectification 
signal discussed above, and to quantitatively compare the
calculated signals with the experimental results.
\begin{figure}
\includegraphics[width=8cm]{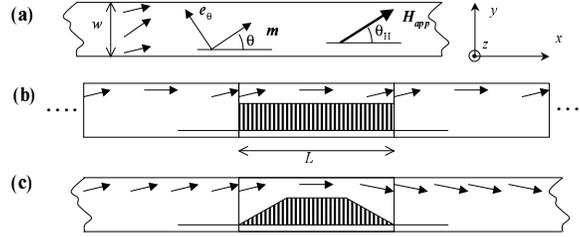}
\caption{ 
\label{fig:geom}
Geometry of the sample and notations definition.
The sample geometry with notations definition is shown in (a).
The two calculation models are schematically depicted: (b) periodic model 
and (c) infinite model.
The hatched area depicts the current $x$ profile.}
\end{figure}

The experimental conditions \cite{Yamaguchi07} are as follows (see 
Fig.~\ref{fig:geom} for notations):
the sample is a magnetic strip, several micrometers long with various
widths (from 300 to 5000~nm) and thicknesses (30 to 50~nm), with the
experimental constraint
of a close to 50~$\Omega$ resistance;
a magnetic field $H_\mathrm{app}$ is applied in the sample plane, at an angle
$\theta_\mathrm{H}$;
an a.c. current with swept frequency is injected into the sample through
a coplanar waveguide; current densities are of the order of 10$^{10}$~A/m$^2$
i.e. low compared to those required for domain wall displacement
\cite{Klaui05a}.
The experimental results may be summarized by:
$(i)$ a d.c. voltage is measured with a marked frequency dependence, it
becomes important ($\approx \mu$V) only at a well defined frequency 
of the order of the ferromagnetic resonance frequency; 
$(ii)$ the position of the resonance is 
strongly influenced by the field magnitude (in accord with Kittel's law);
$(iii)$ the d.c. voltage increases as the square of the 
injected current;
$(iv)$ the angle dependence of the d.c. voltage is well described by a 
$\sin(2\theta_\mathrm{H}) \cos(\theta_\mathrm{H})$ law at large fields, 
turning to $\sin(\theta_\mathrm{H})$ at low fields.

Our approach uses both analytical and numerical micromagnetics: we solve
the Landau-Lifchitz-Gilbert magnetization dynamics equation with incorporation
of the two basic STT terms (\ref{eq:STT}).
Section~\ref{sec:parfait} is devoted to the case of a uniform magnetization 
along the strip axis, where no STT is expected in the linear limit.
The next section introduces structural inhomogeneities, in the spirit of the
well known ripple patterns in thin films \cite{Hubert98}, so that a STT is
present in the ground state.
As both configurations lead to d.c. voltages much below the experimental
levels, Sec.~\ref{sec:field} investigates the effects of a small average 
{\OE}rsted field.
Finally, as the samples have been carefully designed to avoid fields from the 
current 
leads, an intrinsic origin to this field from different electron scattering
properties at the top and bottom surfaces of the strip is discussed.

\section{Uniform magnetization}
\label{sec:parfait}

We first look at the simplest situation where the magnetization does not
change in the direction of the electric field, at least in the rest state
without current.

\subsection{Analytical analysis of the uniform situation}
\label{sec:parf-ana}

The magnetization at rest will be denoted $\vec{m}_0$, with
$| \vec{m}_0 | = 1$.
In the presence of the a.c. current, a small deviation $\vec{m}(\vec{r},t)$
appears ($|\vec{m}| \ll 1$).
The current is described by a spatially uniform $\vec{u}$ that is
harmonic in time with pulsation $\omega$. 
With the axes defined in Fig.~\ref{fig:geom} one has
$\vec{u}=(u(t)=u_0 \cos(\omega t), 0, 0)$.
The LLG equation supplemented by both STT terms reads, to first oder
in the deviations $\vec{m}$
\begin{eqnarray}
\label{eq:LLGparf}
\frac{\partial \vec{m}}{\partial t} &=& 
\gamma_0 \left( \vec{H}_0 \times \vec{m}+ \vec{h} \times \vec{m}_0 \right)
+ \alpha \vec{m}_0 \times \frac{\partial \vec{m}}{\partial t} \nonumber \\
&-& u_0 \cos(\omega t) \frac{\partial \vec{m}}{\partial x}
+ \beta u_0 \cos(\omega t) \vec{m}_0 \times \frac{\partial \vec{m}}{\partial x}.
\end{eqnarray}
In this equation, $\vec{H}_0$ is the effective field of the static magnetization
(with $\vec{H}_0 \times \vec{m}_0 = \vec{0}$ by definition) and $\vec{h}$
is the effective field resulting from the existence of the deviation
$\vec{m}$, with contributions from the exchange and magnetostatic energies.
With only the first two terms on the right-hand side of (\ref{eq:LLGparf}),
upon diagonalization, the various spin wave modes corresponding to the
static magnetization $\vec{m}_0$ are obtained \cite{Bayer06}.
Their amplitude is fixed by the thermal noise.
The last two terms represent the spin-wave pumping by the STT.
We note that they have the form of a product $u(t) \vec{m}$ (we forget the 
spatial derivatives here as the argument is about the time dependence).
Therefore, if $\vec{m}$ varies in time with the pulsation $\omega$ also, 
these terms do not contribute to $\vec{m}$ as they lead to
pulsations $2 \omega$ and $0$.

In fact, a coupling of the form $u(t) \vec{m}$ is known to give rise to
parametric excitation, i.e. the generation of a solution at frequency $f$
by pumping at frequency $2 f$.
Parametric pumping of spin waves from a uniform starting configuration
would not be in agreement with the experimental results (point $(i)$), 
since the resonant frequency found for the current is of the order of the FMR
frequency, not twice this value.

\subsection{Numerical calculations}
\label{sec:parf-num}

As analytical calculations suffer from some limitations, such as linearization
and the consideration of simple structures only, they were completed by
full micromagnetic numerical simulations 
(see Methods in Ref.~\cite{Nakatani03}).
These were performed by solving the LLG equation with the STT terms.
The typical velocity $u_0$ equivalent to the current applied was 
$u_0 = 3.25$~m/s (corresponding to the experimental current density
$J= 6.5 \times 10^{10}$~A/m$^2$ with a polarization $P= 0.7$).
In addition, for the purpose of showing better the effect of STT, values
of $u_0$ as large as 100 or even 1000~m/s were applied.
As no effects of the $\beta$ term were observed (this will become clear
from the analytical calculations), the results shown below were obtained
with $\beta= 0$.
We will throughout the paper show results for just one sample size 
(width $w= 300$~nm and thickness $t= 50$~nm), i.e. values corresponding to 
one sample of Ref.~\cite{Yamaguchi07}.
Material parameters were those representative of Ni$_{80}$Fe$_{20}$,
namely magnetization $M_\mathrm{s}= 800$~kA/m, exchange energy
constant $A= 1 \times 10^{-11}$~J/m, no crystalline anisotropy
(except for the `ripple' case, see Sec.~\ref{sec:ripple}),
gyromagnetic ratio $\gamma_0= 2.21 \times 10^5$~m/(As) and
damping parameter $\alpha= 0.01$.
A static field $H_\mathrm{app}$ was applied in the sample plane,
at an angle $\theta_H$, with standard values 
$\mu_0 H_\mathrm{app}= 40$~mT and $\theta_H=45^\circ$ (the value
where the d.c. signal is close to maximum).
The calculation region length $L$, a part of the real sample length
$L_\mathrm{s}$, was taken to be $L= 1$~$\mu$m, mostly
(calculations with $L= 2$ or 4~$\mu$m were also conducted for the
purpose of checking the dependence on $L$ of the d.c. voltage).
The mesh size was $4 \times 4 \times 50$~nm$^3$ mostly.
The d.c. voltage was computed from the time variation of the
anisotropic magnetoresistance (AMR) of the sample.
Denoting the resistivity change upon magnetization rotation by
$\Delta \rho$ (with, for the used NiFe alloy at room temperature,
$\Delta \rho \approx 0.5 \times 10^{-8}$~$\Omega$m), the dependence of the 
wire resistance upon its magnetization distribution is expressed as
\begin{equation}
\label{eq:DeltaR}
AMR = -\frac{\Delta \rho L_\mathrm{s}}{S} <m_y^2 + m_z^2>,
\end{equation}
where $S$ is the wire cross-section area ($S= w t$) and $<>$ denotes
the average over the calculation region.

Note that, as there is no domain wall in the calculation region and
a field with a transverse component is applied, the calculation
scheme has to be different from that used for the simulation of
domain wall dynamics \cite{Nakatani03}: one cannot assume that
outside the calculation region the magnetization is uniform and
equal to $(\pm 1, 0, 0)$.
Thus, the calculation region was embedded in a wire of infinite length
according to two different models (too simple embedding schemes 
can lead to gradients of $\vec{m}$ in the $x$ direction,
that cause large spurious spin transfer torques).
In the `periodic' model (Fig.~\ref{fig:geom}b), the calculation region is 
supposed to be
repeated periodically in the $x$ direction.
The exchange and demagnetizing fields are calculated accordingly,
as well as the $x$ gradients for the STT.
The current density is uniform.
In the `infinite' model (Fig.~\ref{fig:geom}c), one assumes that the 
values at the left
edge of the calculation region extend to infinity on the left
side, and similarly on the right.
The boundary conditions at left and right are free.
The demagnetizing field takes into account these two semi-inifinite
regions.
In order to avoid end effects, the current density is zero at
$x= 0, L$ and rises (along a length $L/3$) linearly towards the set 
value at the center of the calculation region.

\begin{figure}
\includegraphics[width=8cm]{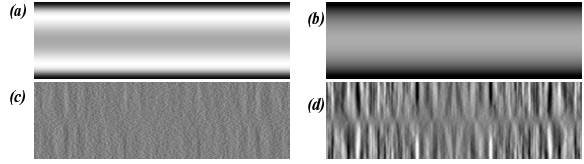}
\caption{ 
\label{fig:parf-init}
Maps of the initial magnetization state under a static field 
$\mu_0 H_\mathrm{app}= 40$~mT ($H_\mathrm{app}= 400$~Oe) applied in the 
film plane at an
angle $\theta_H= 45^\circ$, in a perfect wire 300~nm wide and 50~nm thick.
The axial component $m_x$ is shown in (a), with the gray scale applied
to the deviation of $m_x$ from its average, magnified 50 times.
For the $m_y$ component (b), the gray scale is magnified 20 times 
(from 0.1 to 0.2).
In order to display the magnetization (non) uniformity in $x$ despite the
non uniformity in $y$ due to the inclined applied field, the maps of
the $x$ differential ($m(i+1,j)-m(i-1,j)$, with $i$ and $j$ cell indices) 
of the magnetization components are shown in (c) for $m_x$ and (d) for $m_y$, 
magnified by a factor $10^{15}$.
The images cover the calculation region length (1~$\mu$m).}
\end{figure}

As seen in the analytical analysis (Sec.~\ref{sec:parf-ana}), no effect
is expected in first order perturbation as the initial state magnetization
has no gradient along the current direction.
Numerically however, the avoidance of any $x$ gradient is impossible, as
this would require an infinitely precise numerical evaluation of the
demagnetizing and exchange fields.
It follows that, depending on the model used and the accuracy of the
numerical scheme, a gradient along $x$ of the magnetization remains
that gives rise to a non-zero STT, therefore to a magnetization
oscillation and finally to some d.c. voltage.
The most uniform initial state was obtained with the infinite model, and
is depicted in Fig.~\ref{fig:parf-init} for a field $H_\mathrm{app}$ applied 
at $\theta_H= 45^\circ$.
Note that, as the field is inclined and its value 
$\mu_0 H_\mathrm{app}= 40$~mT
is below the effective transverse anisotropy of the magnetic strip, the 
magnetization becomes non uniform in the transverse $y$ direction, with 
more rotation towards the field at the strip $y$ center.
Fig.~\ref{fig:parf-init} proves that the non uniformity in the $x$
direction of the magnetization is essentially due to the numerical 
precision of the calculations (a double precision number is described
with a relative precision of $10^{-16}$).

\begin{figure}
\includegraphics[width=8cm]{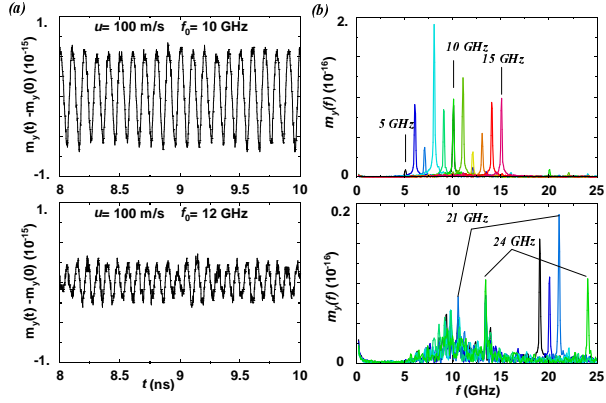}
\caption{ 
\label{fig:parf-m(t)}
(color online) Time variation of the magnetization at one cell in the
calculation region, under a.c. currents of varying frequency and
fixed amplitude $u_0= 100$~m/s, in a perfect sample (to numerical accuracy).
The transverse component $m_y$ is shown, for the sake of sensitivity.
Real time traces are shown in (a), once the steady state oscillation
has been reached, for two frequencies $f_0$ of the a.c. current
(note the $10^{15}$ magnification).
In (b) Fourier spectra are displayed, for lower
frequencies ($f_0= 5$ to 15~GHz with 1~GHz step, top panel) and higher 
frequencies ($f_0= 19$ to 24~GHz with 1~GHz step, bottom panel; note the 
magnified vertical scale).
They reveal the excitation at the fundamental frequency with,
in some frequency range, the apparition of under-multiple peaks
when they correspond to characteristic frequencies of the sample.
The current frequencies $f_0$ corresponding to some spectra are 
indicated.}
\end{figure}

The time variation of the magnetization at position
$(x= L/2, y=w/4)$ in the calculation region, under a.c. current
at various frequencies, is plotted in Fig.~\ref{fig:parf-m(t)} for the
initial state depicted in Fig.~\ref{fig:parf-init}.
The current amplitude was $u_0= 100$~m/s, much higher than the largest
experimental value ($\sim 3$~m/s), but this was necessary in order to
see some effect.
Indeed, the deviations of this central moment are extremely small, of
the order of 10$^{-15}$ whereas the initial value is $m_y= 0.151$ at 
that point.
The Fourier analysis of the data reveals that the fundamental frequency
$f_0$ of the a.c. current
is seen in the spectra only when it is close to a resonant frequency of the 
sample (the numerical calculation of the FMR-like excitation of the
sample by an a.c. field in this configuration shows resonance at
$f_0 \simeq 11$~GHz).
In this regime, multiples of the fundamental frequency are also seen.
In addition, when $f_0$ is close to twice this resonant frequency, 
a (broader) response at half the current frequency can be seen.
This is a characteristic of the parametric excitation,
expected from the analytical analysis presented above.

\begin{figure}
\includegraphics[width=8cm]{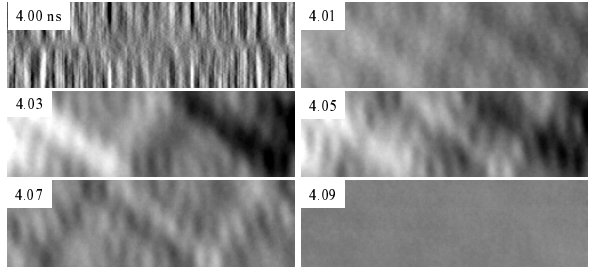}
\caption{ 
\label{fig:parf-movies}
Images of the magnetization oscillating state under a.c. current
at $f_0= 11$~GHz, with $u_0= 100$~m/s, in a perfect sample.
The first image is built from the centered $x$ differential of the
magnetization $y$ component, magnified by a factor $10^{15}$, at 
time $t= 4$~ns after current application (in the steady regime, note
however the closeness to Fig.~\ref{fig:parf-init}d).
The other images are generated by difference of the magnetization
$y$ component with the same component at time $t= 4.0$~ns, magnified by 
$3 \times 10^{14}$. }
\end{figure}

In order to see how this oscillation behaves in space, some snapshots
of the magnetization structure are provided in Fig.~\ref{fig:parf-movies}.
The images reveal that the complex wavy state seen in the structure at rest
(Fig.~\ref{fig:parf-init}) acquires a tiny oscillation in the form of
a quasi-standing wave pattern (with some very geometrical features that
may be artifacts).
Such a pattern had been evidenced in the preliminary calculations
 \cite{Yamaguchi07,Yamaguchi07b}.
However, the amplitude of this oscillation is extremely small 
(in cases where the $x$ gradients of static magnetization at the
left and right edges of the calculation region are less carefully
avoided, the oscillations are stronger).

\begin{figure}
\includegraphics[width=6cm]{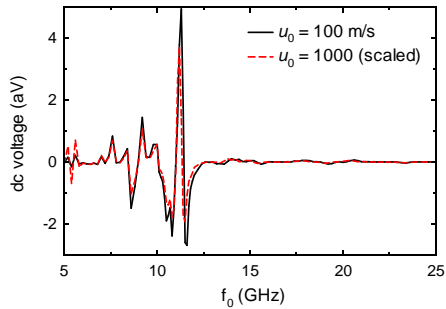}
\caption{ 
\label{fig:parf-dcV}
Spectrum of the d.c. voltage for the uniform initial state.
The current amplitudes are very large, $u_0= 100$ and 1000~m/s.
The values for $u_0= 1000$~m/s are divided by a factor $10^2$ in order to 
compare them to those at $u_0= 100$~m/s.
Note the extreme smallness of the d.c. voltage obtained (vertical
scale in atto-volts ($10^{-18}$~V).}
\end{figure}

Therefore, the computed d.c. voltages are, even at their maximum around
the FMR frequency, orders of magnitude (here, $\approx 15$)  below
the experimental values, as shown in Fig.~\ref{fig:parf-dcV}.
Thus, clearly, another mechanism has to be found.

\section{Non uniform magnetization}
\label{sec:ripple}

The analysis in the preceding section has shown that one of the reasons
for the absence of signal was that the initial state contained no gradient
along the electric field.
A well-known cause for such non uniformitiy is the so-called `ripple'
pattern \cite{Hubert98}, a result of a random distribution of anisotropy
in crystallites.
The response of the magnetization structure to this random potential is to
organize itself in ripple domains, where the magnetization deviates slightly
from its average value in one direction or the other.
These domains have a characteristic lens shape, with the long axis of the lens
oriented normal to the magnetization.
The presence of the ripple structure is attested by Lorentz electron
microscopy (see e.g. Ref.~\cite{Fuller60}).

\subsection{Analytical model for smooth ripple}
\label{sec:ripple-ana}

A first model consists in assuming that the ripple pattern can be
described by a smooth undulation of the magnetization.
Thus we write 
$\vec{m}_0(x,y)=(\cos(\theta_0), \sin(\theta_0), 0)$, where
$\theta_0$ oscillates, more or less regularly, in space.
The small deviation $\vec{m}$ is decomposed into an in-plane 
component and an out-of-plane component:
$\vec{m} = a\vec{e}_{\theta} + b \vec{e}_z$, where 
$\vec{e}_\theta$ is the unit in-plane vector orthogonal to the local
magnetization and $\vec{e}_z$ the unit vector normal to the film plane.
The linearized LLG equation projected on these two vectors now reads
\begin{eqnarray}
\label{eq:LLGripple}
\frac{d a}{d t} &=& \gamma_0 H_b - \alpha \frac{d b}{d t}
- u(t) \frac{\partial \theta_0}{\partial x}, \nonumber \\
\frac{d b}{d t} &=& -\gamma_0 H_a + \alpha \frac{d a}{d t}
+ \beta u(t) \frac{\partial \theta_0}{\partial x}.
\end{eqnarray}
To solve the equation, we assume that the two components of the effective
field are simply proportional to the magnetization deviations :
$H_a = -h_a a$ and $H_b= - h_b b$.
For a thin fim, one expects that $h_b \approx M_\mathrm{s}$ if the ripple
period is much larger than the sample thickness.
The term $h_a$ corresponds to the effective potential that stabilizes
the value of $\theta_0$ and its variation with $x$.
The equations are linear and contain the STT as a driving term now.
They are easily solved in the harmonic approximation :
$a= A \exp(i \omega t)$, $b= B \exp(i \omega t)$ and
$u= U \exp(i \omega t)$.
We then get
\begin{equation}
\label{eq:sol-ripple}
A = U \frac{\partial \theta_0}{\partial x} 
\frac{i\omega + \beta \left( \gamma_0 h_b + \alpha i \omega \right) }
{(1+\alpha^2) \omega^2 -\gamma_0^2 h_a h_b -\alpha i \omega (h_a+h_b)}.
\end{equation}
The magnetization in-plane deviation becomes large at the ferromagnetic
resonance defined by 
$\omega_\mathrm{res}^2 (1+\alpha^2) =\gamma_0^2 h_a h_b$. 
As the meaning of the deviation component $a$ is a rotation of the in-plane 
magnetization angle $\theta_0$, we see that at resonance a current
$u_0 \cos(\omega_\mathrm{res} t)$ transforms the structure as
\begin{equation}
\label{eq:ripple-osc}
\theta_0(x) \rightarrow
\theta_0 \left[x-\frac{u_0 \cos(\omega_\mathrm{res} t)}
{\alpha \gamma_0 (h_a+h_b)} \right].
\end{equation}
This means that the ripple magnetic pattern under STT is set into oscillation
along the $x$ direction.
The amplitude of oscillation at resonance is typically (in the thin film 
approximation where $h_a \ll h_b \approx M_\mathrm{s}$)
\begin{equation}
\label{eq:ampli-ripple}
\Delta x \approx \frac{u_0}{\alpha \gamma_0 M_\mathrm{s}}.
\end{equation}
For $u_0 = 3.25$~m/s, the oscillation amplitude computed from 
(\ref{eq:ampli-ripple}) is $\Delta x = \pm 2$~nm.

The predicted oscillation of the ripple pattern is therefore quite small.
Moreover, in an infinite wire, the modulation of the total
anisotropic magnetoresistance is exactly zero, as the structure is merely 
translated.
For a wire of finite length, some small signal can be expected as, at both
ends, some part of the structure disappears or appears because of the structure
oscillation in position. 
In such situation however, the sign of the d.c. voltage
should be arbitrary and the d.c. voltage
magnitude should not depend on sample length.

\subsection{Numerical calculations}
\label{sec:ripple-num}

\begin{figure}
\includegraphics[width=8cm]{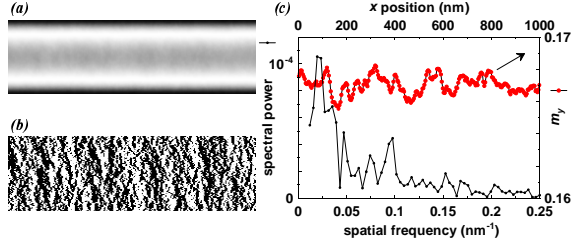}
\caption{ 
\label{fig:ripple-init}
The initial magnetization state for a sample with ripple structure.
The magnetization $x$ component is displayed in (a) with, similarly
to Fig.~\ref{fig:parf-init} a magnification by a factor 50.
In order to better display the magnetization variations in $x$ despite 
the non uniformity in $y$ due to the applied field, the map of the
$x$ differential of $m_y$ is shown in (b), where the centered
cell to cell difference was magnified (slightly over) by a factor $10^{4}$.
The $m_y(x)$ profile averaged over ten lines of cells at the strip
center is shown in (c), with its Fourier transform in order to display
the longitudinal `periodicity' of the ripple pattern.}
\end{figure}

As for the preceding section, numerical calculations were performed
in order to get quantitative results.
For inducing a ripple structure, a random anisotropy field was
introduced, with a fixed value $H_\mathrm{K}$ and a random
in-plane easy axis orientation.
Results obtained for $\mu_0 H_\mathrm{K}= 3$~mT only will be shown here.
The description of the magnetic structure at rest is provided in 
Fig.~\ref{fig:ripple-init}, to be compared to that of the perfect
sample (Fig.~\ref{fig:parf-init}).
It is clear that the random anisotropy leads to much larger magnetization
deviations than the residual ones in the perfect case.
The structure evidences some periodicity along $x$ (about 250~nm, see
a profile of $m_y$ in the $x$ direction together with its Fourier
transform in Fig.~\ref{fig:ripple-init}c).
In the ripple theory \cite{Hubert98}, two structural periods appear, that in 
the direction of magnetization being consequently smaller than that in 
the orthogonal direction.
This larger period appears to be mostly suppressed by the
reduced width of the nanostrip sample, for the parameters chosen here.
Note the similarity of the differential magnetization image in
Fig.~\ref{fig:ripple-init}b with typical TEM images in Lorentz mode, that
also depend on magnetization gradients \cite{Fuller60}.

\begin{figure}
\includegraphics[width=7.3cm]{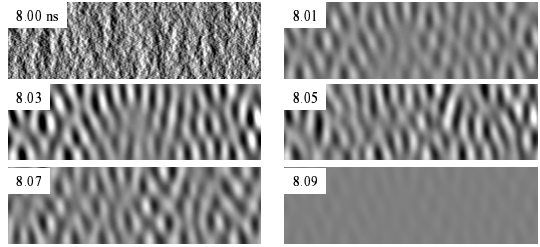}
\caption{ 
\label{fig:ripple-movies}
Images of the magnetization oscillating state under a.c. current
at frequency $f_0= 11$~GHz, with $u_0= 100$~m/s, for
a sample with ripple structure.
The first image, at time $t= 8$~ns, is the centered $x$ differential of the
magnetization $y$ component, magnified by a factor $10^3$.
It is very close to the structure at rest (less magnified).
The next images, corresponding to one period of oscillation, are just
the differences of $m_y(t)$ with $m_y(t=8 \; \mathrm{ns})$, also magnified
by a factor $10^3$.}
\end{figure}

Under application of an a.c. spin polarized current, the magnetization
structure is driven into some oscillation.
Fig.~\ref{fig:ripple-movies} provides some snapshots of the magnetization
distribution, for the high current amplitude $u_0 = 100$~m/s.
Despite the large current, the oscillation nature is hard to apprehend from 
the movies of $m_y(t)$.
Therefore, the time differentials $m_y(t)-m_y(t_0)$ were drawn.
The $y$ component of $\vec{m}$ was chosen as it is the most labile, the
magnetization at rest under the applied field being close to the $x$
direction, in the standard case illustrated here.
A very smooth modulation appears, that contrasts with the noisy appearance
of the rest state gradient image.
It is a standing wave pattern tuned to the current frequency (this pattern
is already discernable in Fig.~\ref{fig:parf-movies} behind the more 
geometrical texture).
Note that the wave pattern is not exactly `standing': the modulation
breathes with time but also deforms slightly during one period.
This deformation looks like the pattern oscillation computed in the
previous subsection, but this requires a more precise verification.
The `wavevector' of this pattern is roughly parallel to the $x$ direction, 
and the wavelength is decreasing as the applied frequency increases (not
shown).
This last feature, compared to the conclusions of the previous subsection,
shows that the dominant mechanism of the spin transfer torque action is
different.

A more appropriate description can be constructed by assuming that 
the ripple pattern is random.
The spin transfer torque due to the $x$ gradients of the structure at rest
$\vec{m}_0$ is equivalent to a field that `pumps' the deviations
$\vec{m}$, expressed as
\begin{equation}
\label{eq:Hripple}
\vec{H}_r(t)=\left[ \left( \frac{\vec{u}(t)}{\gamma_0} 
\cdot \vec{\nabla} \right) \vec{m}_0 \right] \times \vec{m}_0.
\end{equation}
Note that Fig.~\ref{fig:ripple-init}b can also be seen as displaying 
the largest component of this field.
Therefore, in first approximation, this field is random in space but
perfectly harmonic in time.
By Fourier transformation, this field can excite spin wave modes
with a frequency matching the current frequency $f_0$ and a
wavevector contained in the spectrum of $H_r$.
Thus, the field $\vec{H}_r$ is similar to the thermal field that
gives a non zero amplitude to the thermodynamic spin waves (as seen
in Brillouin scattering), with the only difference that instead of
being white in temporal frequency it is monochromatic.
Therefore, the observation of induced spin wave modes is no surprise.
From the analogy with the thermal field we get that the amplitude of
the spin wave is proportional to current (and also to the strength
of the ripple $\partial m_{0y} / \partial x$).

\begin{figure}
\includegraphics[width=7cm]{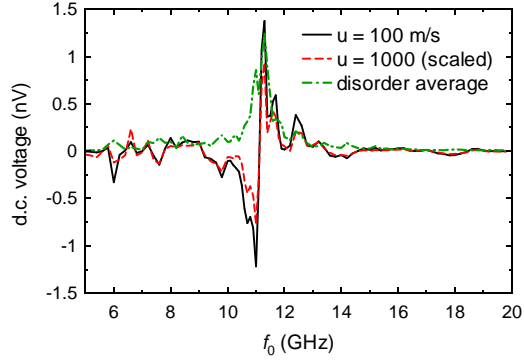}
\caption{ 
\label{fig:ripple-dcV}
Spectrum of the d.c. voltage for one initial state with ripple.
The current amplitude is very large, $u_0= 100$, and the length of the
calculation box is $L= 1$~$\mu$m.
The spectrum obtained with $u_0= 1000$~m/s, divided by a 
factor 100, is superposed (dashed curve).
The dash-dot curve is an average, over 16 realization of a ripple
pattern, of the absolute signals (their sign being random).
}
\end{figure}

The next question concerns the detection of this induced spin wave as
a d.c. voltage.
In first approximation, the wavelength of the standing spin wave
pattern ($\lambda \approx 60$~nm at 11~GHz here) being much smaller
than the length of the sample, the variation of the total AMR
averages to zero.
However, as the standing wave pattern is not perfect, being deformed
by the ripple structure (see Fig.~\ref{fig:ripple-movies}), the 
average is not perfectly zero.
It is a random number with a typical value $\lambda / L$ times smaller 
than the local amplitude.
Fig.~\ref{fig:ripple-dcV} shows a spectrum calculated for one particular
realization of the ripple.
A resonance appears (with a close to expected shape for that
spectrum) at $f_0=11$~GHz, i.e. the FMR frequency of the sample, 
meaning that the uniform mode is also pumped by $\vec{H}_r$.
The values are much larger than in the nominally uniform case
studied earlier (Fig.~\ref{fig:parf-dcV}) and a large irregularity is
present, a signature of the randomness of the averaging over the
sample size.
To this curve is superposed the average of the absolute values of the d.c.
voltages calculated for 16 realizations of the ripple.
Furthermore, a calculation run for a $L=2$~$\mu$m calculation length 
and $u= 100$~m/s gives a similar result, as expected (not shown).
Finally, the voltage is found to be proportional to the square of the 
a.c. current, as expected from the proportionality of spin wave 
amplitude to current
(compare the scaled spectra in Fig.~\ref{fig:ripple-dcV} for 
$u_0= 100$ and $1000$~m/s).
This has the consequence that, for current amplitudes similar to
the experiments, the calculated d.c. voltage should be smaller by a factor
1000, i.e. a million times smaller than what is measured.
Therefore, we conclude that the ripple hypothesis is also not able to
account quantitatively for the measured rectification signal.

\section{Effect of a parasitic field}
\label{sec:field}

The analysis of the first sections has shown that virtually
no rectification effect is to be obtained through a STT as described in
(\ref{eq:STT}), in a single domain ferromagnetic wire.
Of course, we cannot exclude that other mechanisms than those already 
investigated play a role in the interpretation of the experiments, but
we need also to investigate the effect of mechanisms competing with STT.
In the following, we examine the influence of a parasitic {\OE}rsted
field on the rectification effect.

If we consider an infinitely long wire with a uniform current density, the 
average value of the {\OE}rsted field is zero.
As the wire has a flat cross-section, the largest components of the field
are vertical ($z$ direction) and are found at the two lateral edges of the 
wire.
For the $300 \times 50$~nm$^2$ wire submitted to a current density
$J= 6.5 \times 10^{10}$~A/m$^2$, the maximum vertical field
is equal to $\mu_0 H_z = \pm 2.3$~mT (the horizontal components 
reach $\mu_0 H_y = \pm 1.8$~mT). 
These fields, normal to the magnetization at rest, excite magnetization
oscillations in the vicinity of the wire surfaces.
What will be the impact on the sample resistance ?
As a static field is applied in the sample plane, at angle $\theta_H$ from
the wire axis, the $m_y$ component at rest is not zero.
In presence of the {\OE}rsted field, this component oscillates
and therefore the AMR is modulated.
However, as the fields have opposite signs at two opposite surfaces, the
total AMR modulation should cancel by integration.

Note that, in principle, the cancellation may be incomplete.
For example, in the spin-wave community, the effect of interfaces is
often taken into account by a pinning length $\xi$ such that the boundary 
condition for the oscillating magnetization $\vec{m}$ is
\begin{equation}
\label{eq:pinning}
\pm \frac{\partial \vec{m}}{\partial n} + 
\frac{\vec{m}}{\xi}=0,
\end{equation}
where $n$ denotes the coordinate normal to the interface
\cite{Rado59,Guslienko05}.
The original condition \cite{Rado59} specified that 
$\xi = A/ K_\mathrm{s}$, with $K_\mathrm{s}$ the surface
anisotropy constant, but recently magnetization oscillation profile
calculations were shown to be explainable through a pinning length 
related to the sample dimensions \cite{Guslienko05}, actually a
pure magnetostatic effect.
Here, whereas there is no reason to have any dissymetry between opposite
surfaces for magnetostatics, a chemical or structural difference 
between the top and bottom surfaces of the magnetic film cannot
be ruled out, leading to different surface anisotropies.
It is obvious that this will result in a non zero AMR
oscillation and, therefore, a non zero d.c. voltage at the resonance
frequency of this $n=1$ perpendicular standing spin wave mode (PSSW).
Generally, the frequency of the PSSW is different from that of
the uniform FMR mode, allowing their discrimination.
Indeed, on the one hand, frequency should increase due to exchange.
But, on the other hand, the transformation of the lateral magnetostatic 
dynamic charges from monopolar to dipolar decreases the dynamic 
demagnetizing field, thus reducing the frequency.
In the case at hand, the direct numerical calculation gave 10.6~GHz
for the PSSW, close to the value of 11~GHz for the uniform mode,
so that a frequency discrimination is not possible.
Turning now to the signal levels, an upper bound to this contribution
can be obtained by evaluating the d.c. voltage on one half of the
sample thickness.
As peak values of $\pm 1.25$~$\mu$V were obtained, we conclude that 
the d.c. voltage due to a perfect {\OE}rsted field is not sufficient,
by a factor $\sim 10$, to explain the experimental results
Let us therefore now suppose that the average of the {\OE}rsted
field is not exactly zero.

\subsection{Analytical model with a.c. field excitation}
\label{sec:H-ana}

\begin{figure}
\includegraphics[width=8cm]{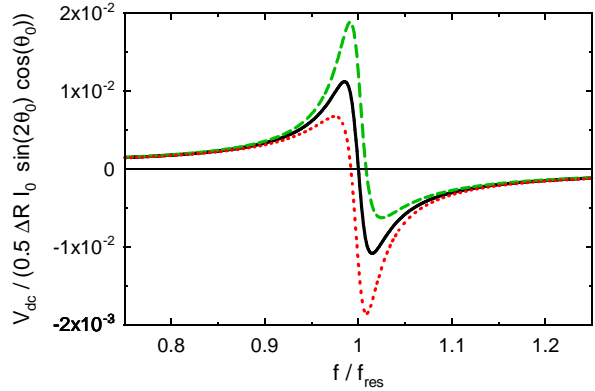}
\caption{ 
\label{fig:Vdc-ana}
Plot of the frequency dependent d.c. voltage according to (\ref{eq:Hsol})
and (\ref{eq:DeltaR(t)}).
The frequency is normalized to resonance, $h_b= M_\mathrm{s}$ and $h_a$ is
such that resonance occurs at 11~GHz, $\alpha= 0.01$, and the magnitude of
the average a.c. field is $\mu_0 H_y= 0.1$~mT.
The dash and dot lines were computed by adding an a.c. field along
the $z$ direction equal to $\pm$ the $y$ field.
This changes the peak shape (less antisymmetric) but not the signal
p.p. amplitude, to first order.
The voltage $\Delta R I_0/2$ is 163~$\mu$V for 
$J_0= 6.5 \times 10^{10}$~A/m$^2$
in the $300 \times 50$~nm$^2$ nanostrip of length 1~$\mu$m,
so that the d.c. signal is calculated to be $\pm 1.4$~$\mu$V. }
\end{figure}

We assume here that the magnetization is uniform (in-plane angle $\theta_0$)
so that from (\ref{eq:DeltaR}) the total AMR is 
$-\Delta \rho L_\mathrm{s} \sin^2(\theta_0) / S$.
When the magnetization angle is oscillating as $\theta_0 + a(t)$, with
$| a(t)| << 1$, the sample resistance
change due to AMR, for a sample of length $L_\mathrm{s}$, reads
\begin{equation}
\label{eq:DeltaR(t)} 
R(t) \approx - \Delta R  \sin(2 \theta_0) a(t).
\end{equation}
where $\Delta R = \Delta \rho L_\mathrm{s} / S$.
We now determine the angle oscillation $a(t)$.
The equations of motion use the same variables as in Sec.~\ref{sec:ripple-ana},
but now instead of considering the STT we include an a.c. field
$h=(0, h_y, h_z)$ (this represents a general non-zero average for the
{\OE}rsted field due to the current, and neglects the zero average part).
The LLG equations of motion now read
\begin{eqnarray}
\label{eq:LLG-H}
\frac{d a}{d t} &=& -\gamma_0 h_b b - \alpha \frac{d b}{d t}
+ \gamma_0 h_z, \nonumber \\
\frac{d b}{d t} &=& \gamma_0 h_a a + \alpha \frac{d a}{d t}
- \gamma_0 h_y \cos(\theta_0).
\end{eqnarray}
The harmonic solution (with $h_y = H_y \exp(i\omega t)$ etc.) reads
similarly
\begin{equation}
\label{eq:Hsol}
A= -\frac{i\omega \gamma_0 H_z + (\gamma_0 h_b + \alpha i \omega)
\gamma_0 H_y \cos(\theta_0)}
{\omega^2 (1+\alpha^2) - \gamma_0^2 h_a h_b -\alpha i \omega \gamma_0
(h_a+h_b)}.
\end{equation}
At resonance, taking into account that $h_b \gg h_a$, the amplitude
is approximately (note that the $y$ field gives the largest effect)
\begin{equation}
\label{eq:ampli-res}
A \approx -i\frac{\gamma_0 H_y \cos(\theta_0)}{\alpha \omega_\mathrm{res}}.
\end{equation}
As this variation of magnetization does not change sign over the sample,
it will not average to zero by integration.
Consequently, this magnetization oscillation produces an oscillation
of the sample resistance at frequency $\omega$, from which a d.c.
voltage results.
The d.c. voltage has a peak close to the resonance frequency : as
the phase of $A$ is close to $\pi/2$ the signal is zero at resonance
but has peaks of alternating signs on both sides (as in the experiments
\cite{Yamaguchi07}).
The typical d.c. voltage is thus
\begin{equation}
\label{eq:Vdc}
V_\mathrm{dc}^\mathrm{typ} = \frac{\Delta R I_0}{2} \sin(2 \theta_0) 
\cos(\theta_0)
\frac{\gamma_0 H_y}{\alpha \omega_\mathrm{res}}.
\end{equation}
With this formula we see that, as in the experiments, the voltage is
proportional to the current square (as the {\OE}rsted field is proportional
to the current), and that the dependence on magnetization angle conforms to
the $\sin(2 \theta) \cos(\theta)$ law.
Fig.~\ref{fig:Vdc-ana} displays the frequency dependent d.c. voltage
evaluated from (\ref{eq:DeltaR(t)}, \ref{eq:Hsol}).
The d.c. voltages are of the order of the micro-Volt, like in the
experiments.
With a sole a.c. field in the $y$ direction, an antisymmetric line
shape is obtained, whereas an a.c. $z$ field contributes with a
symmetric line shape.
Therefore, the experimental line shapes \cite{Yamaguchi07} may be
interpreted by a combination of both of these fields.

We conclude analytically that a non-zero {\OE}rsted field provides a plausible 
explanation of the experimental results showing a rectification effect
in nano and microstrips.
This was qualitatively stated in Ref.~\cite{Yamaguchi07b}, but dismissed
on the ground that the average field should be zero.
The last section of this paper will therefore study
possible origins for such fields.

\subsection{Numerical calculations}
\label{sec:H-num}

The influence of the {\OE}rsted field was investigated by numerical
calculations also.
Indeed, the magnetization non uniformity in the transverse direction,
as a function of applied field angle and magnitude, or the evaluation of
the effect of the full {\OE}rsted field, require a 
micromagnetic simulation.

For the latter point, calculations with a variation along
the $x$ axis are not essential.
Thus, for the description of the full {\OE}rsted field in the sample
cross-section and the calculation of the resulting d.c. voltage, a 2D 
model ($y$, $z$) with invariance in the $x$ direction
was also employed.
The case with no bias (for $h_y$ or $h_z$) was already discussed.
When a bias is added, we find that the d.c. voltage spectra are affected
by the presence of the full {\OE}rsted field only at the frequency of
the PSSW, with a minor quantitative influence when the d.c. voltage is
computed on the full sample thickness.
Therefore, we kept the model used in the rest of the paper for the other
calculations with a.c. field, and neglected the full {\OE}rsted field
whose average is zero.

\begin{figure}
\includegraphics[width=8cm]{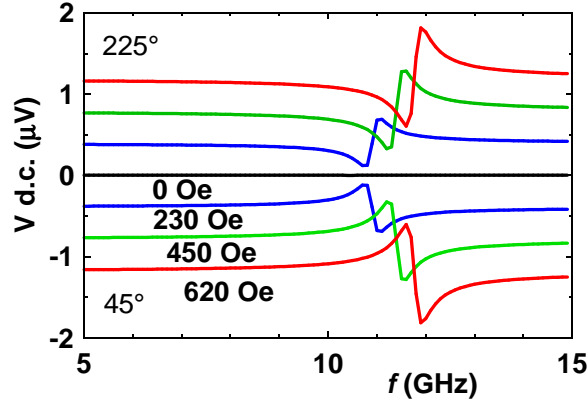}
\caption{ 
\label{fig:Oe-spectra}
Computed d.c. voltage spectra obtained for increasing values of
the applied field, for two field angles $45^\circ$ and $225^\circ$,
and sample dimensions $300 \times 50 \times 1000$~nm$^3$.
An a.c. field of amplitude $\mu_0 H_y= 0.1$~mT is taken into account
as an average {\OE}rsted field,
and the voltage is computed for an a.c. current density 
$J_0= 6.5 \times 10^{10}$~A/m$^2$.
The static field values are chosen to be those of the experiment
\cite{Yamaguchi07}.}
\end{figure}

First, an a.c. field with $y$ component only was considered
$h_y = H_y \cos(\omega t)$ with an amplitude $\mu_0 H_y= 0.1$~mT.
The dependence of the d.c. voltage spectra on the static field
value, for two opposite field directions, is illustrated in
Fig.~\ref{fig:Oe-spectra}.
The field values and angles, as well as the current density, were chosen 
to match those of the experiments \cite{Yamaguchi07}.
In comparison to the analytical calculation assuming uniform magnetization
(Fig.~\ref{fig:Vdc-ana}), the signal is roughly divided by 2.
The results are very similar to experiments, qualitatively and 
quantitatively (the 
computations apply to a calculation region 1~$\mu$m long whereas for these 
dimensions and the nominal resistivity of NiFe the sample length 
corresponding to 50~$\Omega$ is 3~$\mu$m), with 
only a difference in the peak shape.
However, from the analytical modeling (see Fig.~\ref{fig:Vdc-ana}) we
know that such change of shape can be obtained by adding a $z$ component
to the a.c. field.

\begin{figure}
\includegraphics[width=8cm]{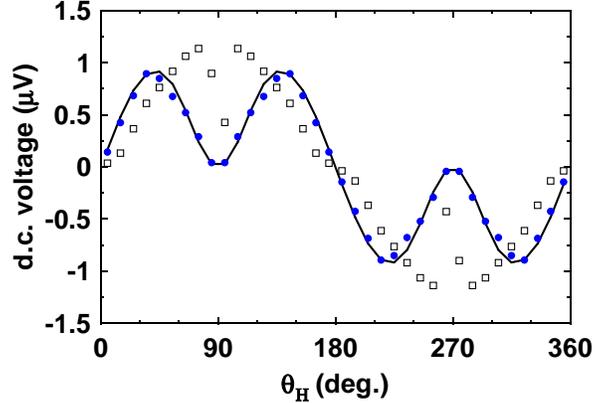}
\caption{ 
\label{fig:Oe-Vdc(angle)}
Computed d.c. voltage maximum amplitude as a function of d.c. field angle.
In order to saturate the nanostrip, a large field has to be applied
($\mu_0 H_\mathrm{app}= 0.5$~T here).
A lower field ($\mu_0 H_\mathrm{app}= 0.16$~T, open squares) leads to a
deformed curve.
Other conditions are identical to those for Fig.~\ref{fig:Oe-spectra}.
The solid line draws the $\cos(2\theta_H)\sin(\theta_H)$ law.}
\end{figure}

Another important experimental feature is the dependence of the
d.c. voltage peak to peak amplitude on the applied field angle. 
It was analytically shown above that the observed behaviour 
(the $\sin(2\theta_\mathrm{H}) \cos(\theta_\mathrm{H})$ law)
was expected for an a.c. $h_y$ field.
The numerical results substantiate this conclusion, as shown
in Fig.~\ref{fig:Oe-Vdc(angle)} for two values of
the static applied field.
For the large value ($\mu_0 H_\mathrm{app}=0.5$~T), the magnetization 
angle $\theta_0$ follows the field angle $\theta_\mathrm{H}$ closely
and the law is well obeyed.
For the smaller field ($\mu_0 H_\mathrm{app}=0.16$~T), closer to the shape
anisotropy field of the nanostrip, an evolution towards a
$\sin(\theta_\mathrm{H})$ variation is evident.

Therefore, assuming that a non zero average {\OE}rsted field
exists, we have been able to reproduce all experimental results,
quantitatively with a.c. fields $\mu_0 H \approx 0.1$~mT.
This is in sharp contrast with the alternative explanations based
on STT, that result in rectification signals that are orders of
magnitude smaller.

\section{Possible origins for a parasitic field}
\label{sec:current}

The question now arises naturally about the origin of such a non zero
average field.
We note that the design of the electrical connections to the magnetic
wire in the experiments \cite{Yamaguchi07} was very symmetrical, the sample
being inserted in a coplanar waveguide with a ground-signal-ground 
structure.
This ensures geometrically that the {\OE}rsted field applied to the 
sample in the presence of a current is minimized.
Of course, any imbalance in current backflow between the 2 ground leads,
or imperfect centering of the sample, creates an a.c. $z$ field 
(for example, we estimate roughly that a 1 \% imbalance would give 
$\mu_0 H_z \sim 0.1$~$\mu$T ).
However, there are two different reasons for which a non-zero average $y$
component of this field can still exist:
(i) the gold coplanar waveguide (100~nm thick) contacts the magnetic sample
by its top surface;
(ii) due to differences in the top and bottom interfaces of the magnetic
film, the current distribution could be non uniform, even in an infinitely 
long magnetic strip.
We try now to estimate these fields.

\subsection{Field from current distribution in the sample thickness}

\begin{figure}
\includegraphics[width=8cm]{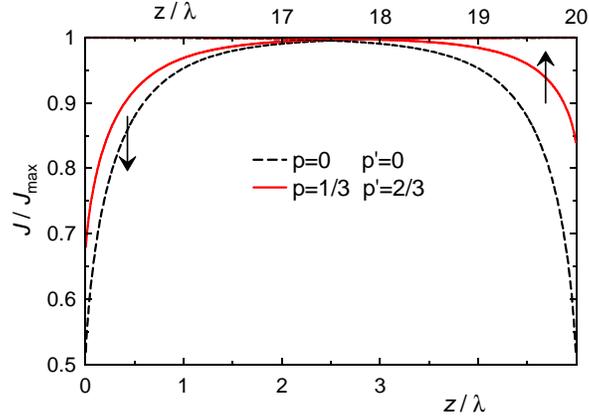}
\caption{ 
\label{fig:J(z)}
Fuchs-Sondheimer model of the current distribution in a film
with thickness $t= 20 \lambda$ where $\lambda$ is the electron mean
free path.
The parameters $p$ and $p'$ are the specularity parameters for the electron
reflection at the bottom and top surfaces, respectively.
The two horizontal scales allow zooming at the vicinity of
the top and bottom surfaces.}
\end{figure}

We first assume that neither a non magnetic conductive underlayer 
nor overlayer exist (that would obviously give rise to a $y$ field),
and look for an intrinsic origin for a non uniform current distribution.
For the description of electronic transport in thin metallic films, the 
Fuchs-Sondheimer model \cite{Sondheimer52} evaluates the
current distribution in the thickness of a thin film, within a 
Boltzmann equation approach.
A key parameter introduced in this model is the specularity parameter
$p$ at every interface: $p=1$ means that electron reflection at the
interface is perfectly specular and $p=0$ that it is random.
In the latter case the current is reduced at the interface.
The characteristic scale (in $z$) over which this reduction extends
is given by the mean free path $\lambda$.
Extending the Fuchs-Sondheimer calculation to different interfaces
at the film top ($z= t$) and bottom ($z= 0$) \cite{Lucas65}, we obtain in the
limit of a thick film ($t \gg \lambda$) the following current
distribution
\begin{equation}
\label{eq:J(z)}
J_x(z) = \sigma_0 E_x \left[ 1 - \frac{1-p}{2} \Phi(\frac{z}{\lambda})
 - \frac{1-p'}{2} \Phi(\frac{t-z}{\lambda}) \right],
\end{equation}
where the function $\Phi$ is an exponential integral
\begin{equation}
\label{eq:Phi}
\Phi(x)= \frac{3}{2} \int_0^1{exp(-\frac{x}{u}) (1-u^2) du}.
\end{equation}
Two $z$ profiles of the current for the case $t= 50$~nm and a mean
free path $\lambda= 2.5$~nm are plotted in Fig.~\ref{fig:J(z)}.
The cases $p=0$, $p'=0$ and $p=1/3$, $p'=2/3$ are compared.
In the former, the current density is reduced to one half at both
interfaces, symetrically.
In the latter, the current is reduced to $2/3$ at one interface
and to $5/6$ at the other, so that $J_x(z)$ becomes asymmetric.

The next step is to evaluate, from this asymmetric current distribution,
the average {\OE}rsted field within the sample.
As $J_x$ is asymmetric in $z$, this gives rise to a non zero average
for the field component transverse to the strip axis ($H_y$).
The analytical calculation of this average field shows that the contribution
of a current layer at depth $z$ to the average field is very close
to linear, the contribution being zero for $z=t/2$.
In other words, the field average over $y$ and $z$ is
\begin{equation}
\label{eq:Hav(J)}
<H_y> = -\frac{2 A}{\pi t} \int_{-t/2}^{t/2}{J_x(z') z' dz'}
\end{equation}
where $A$ is a number equal to $\pi/2$ in the limit $w \gg t$ and 
amounting to $A= 1.296$ for the sample we consider here with $w/t= 6$.
From (\ref{eq:J(z)}) and (\ref{eq:Hav(J)}) we obtain for 
$J= 6.5 \times 10^{10}$~A/m$^2$, $w= 300$~nm and $t= 50$~nm, 
$\mu_0 H_y \approx 5$~$\mu$T.
This value could be perhaps doubled by taking a longer mean free path
and increasingly different specularity parameters, but remains too low 
for explaining the experiments.

\subsection{Field from the contact regions}

In the lack of a full 3D current distribution calculation, the typical field
due to the gold electrodes being deposited on top of the strip surface can be
evaluated roughly.
We assume that, at both ends of the magnetic strip (length $L_\mathrm{s}$), 
the current
flows vertically between the magnetic strip and the gold electrodes, the
latter being twice as thick and 10 times more conductive than the
magnetic sample.
The length of this vertical part is taken to be the thickness $t$ of
the magnetic layer.
At the center of the magnetic strip, considering that 
$L_\mathrm{s} \gg t$, the
resulting typical $y$ field reads
\begin{equation}
\label{eq:Hy-bords}
H_y = \frac{2}{\pi} J w \left( \frac{t}{L_\mathrm{s}} \right) ^2.
\end{equation}
With $J= 6.5 \times 10^{10}$~A/m$^2$, $L_\mathrm{s}=3$~$\mu$m, $w= 300$~nm 
and $t= 50$~nm, one gets $\mu_0 H_y \approx 5$~$\mu$T.
Thus the central value of this field is also too low with regard
to the experiments.
However, it becomes much larger close to the contacts. 
For the average value of the field over the full sample ($x$ and $y$), we obtain
\begin{equation}
\label{eq:Hy-moyen}
<H_y> = \frac{J}{\pi} \frac{t^2}{L_\mathrm{s}} 
\ln( 2 \frac{w}{t})
\end{equation}
(in order to remove a divergence we assume that the current flows 
vertically over a length $t$ along the $x$ axis at the contacts).
The same numbers now give $\mu_0 <H_y>=0.05$~mT,
a value compatible with those explaining the d.c. voltages experimentally
measured.
Therefore, we propose that the rectification signals are due to this field.
It should be noted that this field decreases as the inverse of the sample length, 
in contrast to
the intrinsic mechanism whose the contribution is independent of sample length.

\section{Conclusion}

This work has tried to reach a quantitative description of the
spectra of d.c. voltage versus frequency obtained on nanostrips 
subjected to a static in plane field.
We have found that, for a perfect wire, no signal should be
expected owing to a linear analysis.
The consideration of internal magnetic inhomogeneities in the
nanostrip (ripple structure) do lead to non zero rectification voltages.
However, these are disorder dependent and much too low compared
to experimental findings.
As a solution to this paradox, we tested the influence of a
non zero average value of the {\OE}rsted field generated by
the a.c. current, and found that small values (0.1~mT) of this
average lead to quantitative agreement with experimental findings.
We propose two origins for these fields.
The extrinsic one is
due to the contacts put on the magnetic sample (as considered
recently in the case of vortex excitation by a.c. currents
\cite{Bolte08}).
The intrinsic one is due to an asymmetric current distribution
in the thickness of the magnetic nanostrip.
In the semi-classical description by Fuchs and Sondheimer of
the conductivity of very thin films, such an asymmetry results
from different specularity parameters for electron reflection
at the top and bottom surfaces of the magnetic layer.

Therefore, we conclude that {\OE}rsted field effects should
be carefully investigated when interpreting experiments by
the spin transfer torque mechanism.
The extrinsic contribution should be precisely evaluated, so that
the numerical evaluation of the field taking into account the
full sample structure may become mandantory in the future.
The important role of small a.c. fields at resonant frequencies is
not so surprising in fact: in ferromagnetic resonance, for
a sample with $\alpha=0.01$ like NiFe, the excitation field is
of the order of $\mu_0 H= 1$~$\mu$T.

\section{Acknowledgements}

Comments by J. Miltat and Y. Suzuki are gratefully acknowledged.
The research of A.T. was supported by the french ANR PNANO programme,
(DYNAWALL project), the Programme pluriformations SPINEL of 
Universit{\'{e}} Paris-sud, and the european network MRTN-CT-2006-035327 
SPINSWITCH.
The work of Y.N. was supported partly by a grant-in-aid for scientific 
research in the priority area ``Creation and control of spin current'' from 
the Ministry of Education, Culture, Sports, Science and Technology, Japan.

Note added in proof : A. Yamaguchi, in a recent work \cite{Yamaguchi08}, 
comes also to the conclusion that a field effect is very probably the 
origin of the rectification signal.
In addition, as source of the non zero average field, he proposes the
deformation of the field pattern at high frequencies due to
electromagnetic effects, the permittivity of the substrate being 
different from that of air.

\bibliography{STT,DWdyn,mag-basics,mumag}

\begin{thebibliography}{28}
\expandafter\ifx\csname natexlab\endcsname\relax\def\natexlab#1{#1}\fi
\expandafter\ifx\csname bibnamefont\endcsname\relax
  \def\bibnamefont#1{#1}\fi
\expandafter\ifx\csname bibfnamefont\endcsname\relax
  \def\bibfnamefont#1{#1}\fi
\expandafter\ifx\csname citenamefont\endcsname\relax
  \def\citenamefont#1{#1}\fi
\expandafter\ifx\csname url\endcsname\relax
  \def\url#1{\texttt{#1}}\fi
\expandafter\ifx\csname urlprefix\endcsname\relax\def\urlprefix{URL }\fi
\providecommand{\bibinfo}[2]{#2}
\providecommand{\eprint}[2][]{\url{#2}}

\bibitem[{\citenamefont{Berger}(1996)}]{Berger96}
\bibinfo{author}{\bibfnamefont{L.}~\bibnamefont{Berger}},
  \bibinfo{journal}{Phys.\ Rev.\ B} \textbf{\bibinfo{volume}{54}},
  \bibinfo{pages}{9353} (\bibinfo{year}{1996}).

\bibitem[{\citenamefont{Slonczewski}(1996)}]{Slonczewski96}
\bibinfo{author}{\bibfnamefont{J.}~\bibnamefont{Slonczewski}},
  \bibinfo{journal}{J.\ Magn.\ Magn.\ Mater.} \textbf{\bibinfo{volume}{159}},
  \bibinfo{pages}{L1} (\bibinfo{year}{1996}).

\bibitem[{\citenamefont{Krivorotov et~al.}(2005)\citenamefont{Krivorotov,
  Emley, Sankey, Kiselev, Ralph, and Buhrman}}]{Krivorotov05}
\bibinfo{author}{\bibfnamefont{I.}~\bibnamefont{Krivorotov}},
  \bibinfo{author}{\bibfnamefont{N.}~\bibnamefont{Emley}},
  \bibinfo{author}{\bibfnamefont{J.}~\bibnamefont{Sankey}},
  \bibinfo{author}{\bibfnamefont{S.}~\bibnamefont{Kiselev}},
  \bibinfo{author}{\bibfnamefont{D.}~\bibnamefont{Ralph}}, \bibnamefont{and}
  \bibinfo{author}{\bibfnamefont{R.}~\bibnamefont{Buhrman}},
  \bibinfo{journal}{Science} \textbf{\bibinfo{volume}{307}},
  \bibinfo{pages}{228} (\bibinfo{year}{2005}).

\bibitem[{\citenamefont{Berkov and Miltat}(2008)}]{Berkov08}
\bibinfo{author}{\bibfnamefont{D.}~\bibnamefont{Berkov}} \bibnamefont{and}
  \bibinfo{author}{\bibfnamefont{J.}~\bibnamefont{Miltat}},
  \bibinfo{journal}{J.\ Magn.\ Magn.\ Mater.} \textbf{\bibinfo{volume}{320}},
  \bibinfo{pages}{1238} (\bibinfo{year}{2008}).

\bibitem[{\citenamefont{Bazaliy et~al.}(1998)\citenamefont{Bazaliy, Jones, and
  Zhang}}]{Bazaliy98}
\bibinfo{author}{\bibfnamefont{Y.~B.} \bibnamefont{Bazaliy}},
  \bibinfo{author}{\bibfnamefont{B.}~\bibnamefont{Jones}}, \bibnamefont{and}
  \bibinfo{author}{\bibfnamefont{S.-C.} \bibnamefont{Zhang}},
  \bibinfo{journal}{Phys.\ Rev.\ B} \textbf{\bibinfo{volume}{57}},
  \bibinfo{pages}{R3213} (\bibinfo{year}{1998}).

\bibitem[{\citenamefont{Thiaville et~al.}(2004)\citenamefont{Thiaville,
  Nakatani, Miltat, and Vernier}}]{Thiaville04}
\bibinfo{author}{\bibfnamefont{A.}~\bibnamefont{Thiaville}},
  \bibinfo{author}{\bibfnamefont{Y.}~\bibnamefont{Nakatani}},
  \bibinfo{author}{\bibfnamefont{J.}~\bibnamefont{Miltat}}, \bibnamefont{and}
  \bibinfo{author}{\bibfnamefont{N.}~\bibnamefont{Vernier}},
  \bibinfo{journal}{J.\ Appl.\ Phys.} \textbf{\bibinfo{volume}{95}},
  \bibinfo{pages}{7049} (\bibinfo{year}{2004}).

\bibitem[{\citenamefont{Zhang and Li}(2004)}]{Zhang04}
\bibinfo{author}{\bibfnamefont{S.}~\bibnamefont{Zhang}} \bibnamefont{and}
  \bibinfo{author}{\bibfnamefont{Z.}~\bibnamefont{Li}},
  \bibinfo{journal}{Phys.\ Rev.\ Lett.} \textbf{\bibinfo{volume}{93}},
  \bibinfo{pages}{127204} (\bibinfo{year}{2004}).

\bibitem[{\citenamefont{Thiaville et~al.}(2005)\citenamefont{Thiaville,
  Nakatani, Miltat, and Suzuki}}]{Thiaville05}
\bibinfo{author}{\bibfnamefont{A.}~\bibnamefont{Thiaville}},
  \bibinfo{author}{\bibfnamefont{Y.}~\bibnamefont{Nakatani}},
  \bibinfo{author}{\bibfnamefont{J.}~\bibnamefont{Miltat}}, \bibnamefont{and}
  \bibinfo{author}{\bibfnamefont{Y.}~\bibnamefont{Suzuki}},
  \bibinfo{journal}{Europhys.\ Lett.} \textbf{\bibinfo{volume}{69}},
  \bibinfo{pages}{990} (\bibinfo{year}{2005}).

\bibitem[{\citenamefont{Stiles et~al.}(2007)\citenamefont{Stiles, Saslow,
  Donahue, and Zangwill}}]{Stiles07}
\bibinfo{author}{\bibfnamefont{M.}~\bibnamefont{Stiles}},
  \bibinfo{author}{\bibfnamefont{W.}~\bibnamefont{Saslow}},
  \bibinfo{author}{\bibfnamefont{M.}~\bibnamefont{Donahue}}, \bibnamefont{and}
  \bibinfo{author}{\bibfnamefont{A.}~\bibnamefont{Zangwill}},
  \bibinfo{journal}{Phys.\ Rev.\ B} \textbf{\bibinfo{volume}{75}},
  \bibinfo{pages}{214423} (\bibinfo{year}{2007}).

\bibitem[{\citenamefont{Smith}()}]{Smith07}
\bibinfo{author}{\bibfnamefont{N.}~\bibnamefont{Smith}},
  \bibinfo{note}{cond-mat/0706.1736}.

\bibitem[{\citenamefont{Yamaguchi et~al.}(2007)\citenamefont{Yamaguchi,
  Miyajima, Ono, Suzuki, Yuasa, Tulapurkar, and Nakatani}}]{Yamaguchi07}
\bibinfo{author}{\bibfnamefont{A.}~\bibnamefont{Yamaguchi}},
  \bibinfo{author}{\bibfnamefont{H.}~\bibnamefont{Miyajima}},
  \bibinfo{author}{\bibfnamefont{T.}~\bibnamefont{Ono}},
  \bibinfo{author}{\bibfnamefont{Y.}~\bibnamefont{Suzuki}},
  \bibinfo{author}{\bibfnamefont{S.}~\bibnamefont{Yuasa}},
  \bibinfo{author}{\bibfnamefont{A.}~\bibnamefont{Tulapurkar}},
  \bibnamefont{and} \bibinfo{author}{\bibfnamefont{Y.}~\bibnamefont{Nakatani}},
  \bibinfo{journal}{Appl.\ Phys.\ Lett.} \textbf{\bibinfo{volume}{90}},
  \bibinfo{pages}{182507} (\bibinfo{year}{2007}).

\bibitem[{\citenamefont{Bedau et~al.}(2007)\citenamefont{Bedau, Kl{{\"a}}ui,
  Krzyk, R{{\"u}}diger, Faini, and Vila}}]{Bedau07}
\bibinfo{author}{\bibfnamefont{D.}~\bibnamefont{Bedau}},
  \bibinfo{author}{\bibfnamefont{M.}~\bibnamefont{Kl{{\"a}}ui}},
  \bibinfo{author}{\bibfnamefont{S.}~\bibnamefont{Krzyk}},
  \bibinfo{author}{\bibfnamefont{U.}~\bibnamefont{R{{\"u}}diger}},
  \bibinfo{author}{\bibfnamefont{G.}~\bibnamefont{Faini}}, \bibnamefont{and}
  \bibinfo{author}{\bibfnamefont{L.}~\bibnamefont{Vila}},
  \bibinfo{journal}{Phys.\ Rev.\ Lett.} \textbf{\bibinfo{volume}{99}},
  \bibinfo{pages}{146601} (\bibinfo{year}{2007}).

\bibitem[{\citenamefont{Moriya et~al.}(2008)\citenamefont{Moriya, Thomas,
  Hayashi, Bazaliy, Rettner, and Parkin}}]{Moriya08}
\bibinfo{author}{\bibfnamefont{R.}~\bibnamefont{Moriya}},
  \bibinfo{author}{\bibfnamefont{L.}~\bibnamefont{Thomas}},
  \bibinfo{author}{\bibfnamefont{M.}~\bibnamefont{Hayashi}},
  \bibinfo{author}{\bibfnamefont{Y.}~\bibnamefont{Bazaliy}},
  \bibinfo{author}{\bibfnamefont{C.}~\bibnamefont{Rettner}}, \bibnamefont{and}
  \bibinfo{author}{\bibfnamefont{S.}~\bibnamefont{Parkin}},
  \bibinfo{journal}{Nature Phys.} \textbf{\bibinfo{volume}{4}},
  \bibinfo{pages}{368} (\bibinfo{year}{2008}).

\bibitem[{\citenamefont{Tatara et~al.}(2007)\citenamefont{Tatara, Kohno,
  Shibata, Lemaho, and Lee}}]{Tatara07}
\bibinfo{author}{\bibfnamefont{G.}~\bibnamefont{Tatara}},
  \bibinfo{author}{\bibfnamefont{H.}~\bibnamefont{Kohno}},
  \bibinfo{author}{\bibfnamefont{J.}~\bibnamefont{Shibata}},
  \bibinfo{author}{\bibfnamefont{Y.}~\bibnamefont{Lemaho}}, \bibnamefont{and}
  \bibinfo{author}{\bibfnamefont{K.-J.} \bibnamefont{Lee}},
  \bibinfo{journal}{J.\ Phys.\ Soc.\ Jpn.} \textbf{\bibinfo{volume}{76}},
  \bibinfo{pages}{054707} (\bibinfo{year}{2007}).

\bibitem[{\citenamefont{Pi{\'{e}}chon and Thiaville}(2007)}]{Piechon07}
\bibinfo{author}{\bibfnamefont{F.}~\bibnamefont{Pi{\'{e}}chon}}
  \bibnamefont{and}
  \bibinfo{author}{\bibfnamefont{A.}~\bibnamefont{Thiaville}},
  \bibinfo{journal}{Phys.\ Rev.\ B} \textbf{\bibinfo{volume}{75}},
  \bibinfo{pages}{174414} (\bibinfo{year}{2007}).

\bibitem[{\citenamefont{Shibata et~al.}(2005)\citenamefont{Shibata, Tatara, and
  Kohno}}]{Shibata05}
\bibinfo{author}{\bibfnamefont{J.}~\bibnamefont{Shibata}},
  \bibinfo{author}{\bibfnamefont{G.}~\bibnamefont{Tatara}}, \bibnamefont{and}
  \bibinfo{author}{\bibfnamefont{H.}~\bibnamefont{Kohno}},
  \bibinfo{journal}{Phys.\ Rev.\ Lett.} \textbf{\bibinfo{volume}{94}},
  \bibinfo{pages}{076601} (\bibinfo{year}{2005}).

\bibitem[{\citenamefont{Kl{{\"a}}ui et~al.}(2005)\citenamefont{Kl{{\"a}}ui,
  Vaz, Bland, Wernsdorfer, Faini, Cambril, Heyderman, Nolting, and
  R{{\"u}}diger}}]{Klaui05a}
\bibinfo{author}{\bibfnamefont{M.}~\bibnamefont{Kl{{\"a}}ui}},
  \bibinfo{author}{\bibfnamefont{C.}~\bibnamefont{Vaz}},
  \bibinfo{author}{\bibfnamefont{J.}~\bibnamefont{Bland}},
  \bibinfo{author}{\bibfnamefont{W.}~\bibnamefont{Wernsdorfer}},
  \bibinfo{author}{\bibfnamefont{G.}~\bibnamefont{Faini}},
  \bibinfo{author}{\bibfnamefont{E.}~\bibnamefont{Cambril}},
  \bibinfo{author}{\bibfnamefont{L.}~\bibnamefont{Heyderman}},
  \bibinfo{author}{\bibfnamefont{F.}~\bibnamefont{Nolting}}, \bibnamefont{and}
  \bibinfo{author}{\bibfnamefont{U.}~\bibnamefont{R{{\"u}}diger}},
  \bibinfo{journal}{Phys.\ Rev.\ Lett.} \textbf{\bibinfo{volume}{94}},
  \bibinfo{pages}{106601} (\bibinfo{year}{2005}).

\bibitem[{\citenamefont{Hubert and Sch{{\"a}}fer}(1998)}]{Hubert98}
\bibinfo{author}{\bibfnamefont{A.}~\bibnamefont{Hubert}} \bibnamefont{and}
  \bibinfo{author}{\bibfnamefont{R.}~\bibnamefont{Sch{{\"a}}fer}},
  \emph{\bibinfo{title}{Magnetic Domains}} (\bibinfo{publisher}{Springer
  Verlag}, \bibinfo{address}{Berlin}, \bibinfo{year}{1998}).

\bibitem[{\citenamefont{Bayer et~al.}(2006)\citenamefont{Bayer, Jorzick,
  Demokritov, Slavin, Guslienko, Berkov, Gorn, Kostylev, and
  Hillebrands}}]{Bayer06}
\bibinfo{author}{\bibfnamefont{C.}~\bibnamefont{Bayer}},
  \bibinfo{author}{\bibfnamefont{J.}~\bibnamefont{Jorzick}},
  \bibinfo{author}{\bibfnamefont{S.}~\bibnamefont{Demokritov}},
  \bibinfo{author}{\bibfnamefont{A.}~\bibnamefont{Slavin}},
  \bibinfo{author}{\bibfnamefont{K.}~\bibnamefont{Guslienko}},
  \bibinfo{author}{\bibfnamefont{D.}~\bibnamefont{Berkov}},
  \bibinfo{author}{\bibfnamefont{N.}~\bibnamefont{Gorn}},
  \bibinfo{author}{\bibfnamefont{M.}~\bibnamefont{Kostylev}}, \bibnamefont{and}
  \bibinfo{author}{\bibfnamefont{B.}~\bibnamefont{Hillebrands}},
  \emph{\bibinfo{title}{Spin Dynamics in Confined Magnetic Structures III}}
  (\bibinfo{publisher}{Springer}, \bibinfo{address}{Berlin},
  \bibinfo{year}{2006}), pp. \bibinfo{pages}{57--104}.

\bibitem[{\citenamefont{Nakatani et~al.}(2003)\citenamefont{Nakatani,
  Thiaville, and Miltat}}]{Nakatani03}
\bibinfo{author}{\bibfnamefont{Y.}~\bibnamefont{Nakatani}},
  \bibinfo{author}{\bibfnamefont{A.}~\bibnamefont{Thiaville}},
  \bibnamefont{and} \bibinfo{author}{\bibfnamefont{J.}~\bibnamefont{Miltat}},
  \bibinfo{journal}{Nature Mater.} \textbf{\bibinfo{volume}{2}},
  \bibinfo{pages}{521} (\bibinfo{year}{2003}).

\bibitem[{\citenamefont{Yamaguchi et~al.}()\citenamefont{Yamaguchi, Motoi,
  Miyajima, and Nakatani}}]{Yamaguchi07b}
\bibinfo{author}{\bibfnamefont{A.}~\bibnamefont{Yamaguchi}},
  \bibinfo{author}{\bibfnamefont{K.}~\bibnamefont{Motoi}},
  \bibinfo{author}{\bibfnamefont{H.}~\bibnamefont{Miyajima}}, \bibnamefont{and}
  \bibinfo{author}{\bibfnamefont{Y.}~\bibnamefont{Nakatani}},
  \bibinfo{note}{cond-mat/0710.2172}.

\bibitem[{\citenamefont{Fuller and Hale}(1960)}]{Fuller60}
\bibinfo{author}{\bibfnamefont{H.}~\bibnamefont{Fuller}} \bibnamefont{and}
  \bibinfo{author}{\bibfnamefont{M.}~\bibnamefont{Hale}}, \bibinfo{journal}{J.\
  Appl.\ Phys.} \textbf{\bibinfo{volume}{31}}, \bibinfo{pages}{238}
  (\bibinfo{year}{1960}).

\bibitem[{\citenamefont{Rado and Weertman}(1959)}]{Rado59}
\bibinfo{author}{\bibfnamefont{G.}~\bibnamefont{Rado}} \bibnamefont{and}
  \bibinfo{author}{\bibfnamefont{J.}~\bibnamefont{Weertman}},
  \bibinfo{journal}{J.\ Phys.\ Chem.\ Solids} \textbf{\bibinfo{volume}{11}},
  \bibinfo{pages}{315} (\bibinfo{year}{1959}).

\bibitem[{\citenamefont{Guslienko and Slavin}(2005)}]{Guslienko05}
\bibinfo{author}{\bibfnamefont{K.}~\bibnamefont{Guslienko}} \bibnamefont{and}
  \bibinfo{author}{\bibfnamefont{A.}~\bibnamefont{Slavin}},
  \bibinfo{journal}{Phys.\ Rev.\ B} \textbf{\bibinfo{volume}{72}},
  \bibinfo{pages}{014463} (\bibinfo{year}{2005}).

\bibitem[{\citenamefont{Sondheimer}(1952)}]{Sondheimer52}
\bibinfo{author}{\bibfnamefont{E.}~\bibnamefont{Sondheimer}},
  \bibinfo{journal}{Adv.\ Phy.} \textbf{\bibinfo{volume}{1}},
  \bibinfo{pages}{1} (\bibinfo{year}{1952}).

\bibitem[{\citenamefont{Lucas}(1965)}]{Lucas65}
\bibinfo{author}{\bibfnamefont{M.}~\bibnamefont{Lucas}}, \bibinfo{journal}{J.\
  Appl.\ Phys.} \textbf{\bibinfo{volume}{36}}, \bibinfo{pages}{1632}
  (\bibinfo{year}{1965}).

\bibitem[{\citenamefont{Bolte et~al.}(2008)\citenamefont{Bolte, Meier,
  Kr{\"{u}}ger, Drews, Eiselt, Bocklage, Bohlens, Tyliszczak, Vansteenkiste,
  Waeyenberge et~al.}}]{Bolte08}
\bibinfo{author}{\bibfnamefont{M.}~\bibnamefont{Bolte}},
  \bibinfo{author}{\bibfnamefont{G.}~\bibnamefont{Meier}},
  \bibinfo{author}{\bibfnamefont{B.}~\bibnamefont{Kr{\"{u}}ger}},
  \bibinfo{author}{\bibfnamefont{A.}~\bibnamefont{Drews}},
  \bibinfo{author}{\bibfnamefont{R.}~\bibnamefont{Eiselt}},
  \bibinfo{author}{\bibfnamefont{L.}~\bibnamefont{Bocklage}},
  \bibinfo{author}{\bibfnamefont{S.}~\bibnamefont{Bohlens}},
  \bibinfo{author}{\bibfnamefont{T.}~\bibnamefont{Tyliszczak}},
  \bibinfo{author}{\bibfnamefont{A.}~\bibnamefont{Vansteenkiste}},
  \bibinfo{author}{\bibfnamefont{B.~V.} \bibnamefont{Waeyenberge}},
  \bibnamefont{et~al.}, \bibinfo{journal}{Phys.\ Rev.\ Lett.}
  \textbf{\bibinfo{volume}{100}}, \bibinfo{pages}{176601}
  (\bibinfo{year}{2008}).

\bibitem[{\citenamefont{Yamaguchi et~al.}(2008)\citenamefont{Yamaguchi, Motoi,
  Hirohata, Miyajima, Miyashita, and Sanada}}]{Yamaguchi08}
\bibinfo{author}{\bibfnamefont{A.}~\bibnamefont{Yamaguchi}},
  \bibinfo{author}{\bibfnamefont{K.}~\bibnamefont{Motoi}},
  \bibinfo{author}{\bibfnamefont{A.}~\bibnamefont{Hirohata}},
  \bibinfo{author}{\bibfnamefont{H.}~\bibnamefont{Miyajima}},
  \bibinfo{author}{\bibfnamefont{Y.}~\bibnamefont{Miyashita}},
  \bibnamefont{and} \bibinfo{author}{\bibfnamefont{Y.}~\bibnamefont{Sanada}},
  \bibinfo{journal}{Phys.\ Rev.\ B} \textbf{\bibinfo{volume}{78}},
  \bibinfo{pages}{104401} (\bibinfo{year}{2008}).

\end{thebibliography}

\end{document}